\documentclass[conference]{IEEEtran}
\usepackage{cite}
\usepackage{amsmath,amssymb,amsfonts}
\usepackage{algorithmic}
\usepackage{graphicx}
\usepackage{textcomp}
\usepackage{xcolor}
\usepackage[hyphens]{url}
\usepackage{hyperref}

\usepackage{balance}

\def\BibTeX{{\rm B\kern-.05em{\sc i\kern-.025em b}\kern-.08em
    T\kern-.1667em\lower.7ex\hbox{E}\kern-.125emX}}

\title{Towards Autonomous Accelerator Design: \\ FPGA Accelerator Generation with SECDA}

\author{%
\textbf{Vinamra Sharma}\quad
\textbf{Xingjian Fu}\quad
\textbf{Jude Haris}\quad
\textbf{Jos\'e Cano }$^{\quad}$\\
School of Computing Science, University of Glasgow, Scotland, UK\quad
\\
\footnotesize{\texttt{{\{\href{mailto:2805343s@student.gla.ac.uk}{2805343s},\,\href{mailto:3065487f@student.gla.ac.uk}{3065487f}\}@student.gla.ac.uk}}}\,,\footnotesize{\texttt{\{\href{mailto:Jude.Haris@glasgow.ac.uk}{jude.haris},\,\href{mailto:jose.canoreyes@glasgow.ac.uk}{jose.canoreyes},\}@glasgow.ac.uk}}
}
\usepackage{fancyhdr}
\fancypagestyle{firstpage}{
  \fancyhf{} 
  \chead{\small \textit{Accepted to the Machine Learning for Architecture and Systems Workshop (\href{https://sites.google.com/corp/view/mlarchsys}{MLArchSys}), co-located with \href{https://iscaconf.org/isca2026/}{ISCA 2026}.}}
  \cfoot{\thepage} 
}

\begin{document}
\maketitle

\thispagestyle{firstpage} 
\pagestyle{plain}



\begin{abstract}

Designing FPGA-based accelerators for modern artificial intelligence workloads requires exploring a large and complex hardware design space that involves architectural parameters, data flow strategies, and memory hierarchies, making the process very time consuming. 
While existing methodologies such as SECDA enable rapid hardware-software co-design through SystemC simulation and FPGA execution, identifying efficient accelerator configurations remains a largely manual process requiring extensive domain knowledge. 
SECDA-DSE is a framework that integrates Large Language Models (LLMs) into the SECDA ecosystem to guide design space exploration (DSE) of FPGA-based accelerators. It combines a structured DSE Explorer for generating candidate architectures with an LLM Stack that performs reasoning-guided exploration using retrieval-augmented generation and chain-of-thought prompting, coupled with a feedback loop for iterative and reinforced refinement.

Building on our previous work introducing SECDA-DSE, this paper extends its evaluation by generating three accelerator designs, including element-wise vector multiplication, 2D convolution, and matrix transpose, and performing end-to-end execution on FPGA hardware.
The results show that SECDA-DSE can generate SECDA-compliant accelerator designs that are successfully synthesized and executed on FPGA hardware. Furthermore, the generated designs capture kernel-specific trade-offs between compute parallelism and data movement, highlighting the potential of LLM-guided exploration to adapt architectural configurations across diverse workloads while reducing exploration time and the need for extensive human expertise.

\end{abstract}

\section{Introduction}
\label{intro}


Designing FPGA-based accelerators for modern artificial intelligence (AI) workloads requires selecting among interacting architectural choices, including compute parallelism, tiling factors, memory hierarchy organization, dataflow strategies, and on-chip resource allocation. These choices jointly affect latency, throughput, bandwidth pressure, and FPGA resource utilization, making accelerator design space exploration both expensive and expertise-intensive~\cite{mohaidat2024survey}. 
Although FPGA-based accelerators provide flexibility and energy efficiency compared to fixed-function ASICs, identifying performant accelerator configurations remains a significant challenge due to the large number of architectural permutations and optimization trade-offs.

Existing methodologies such as SECDA~\cite{haris2021secda} and toolkits like SECDA-TFLite~\cite{haris2023secda} enable rapid hardware software co-design through SystemC-based simulation and FPGA execution, significantly reducing the development overhead associated with custom accelerator design. They provide reusable architectural templates and automated hardware flows for AI acceleration. However, the process of exploring optimal accelerator configurations still requires extensive manual design space exploration (DSE), iterative tuning, and expert-driven reasoning over hardware trade-offs.

Recent advances in Large Language Models (LLMs) have demonstrated strong capabilities in reasoning, code generation, and structured problem solving across software engineering and systems domains~\cite{hou2024large, fan2023large}. These capabilities create an opportunity to assist or automate hardware accelerator exploration by enabling reasoning-guided design refinement and workload-aware architectural adaptation. However, applying LLMs to FPGA accelerator DSE remains relatively unexplored, particularly in combining LLM reasoning with structured hardware exploration flows and validating generated accelerator designs through real FPGA execution.

In our previous work~\cite{sharma2026llm}, we introduced \textbf{SECDA-DSE}, a framework that integrates LLM-guided reasoning into the SECDA ecosystem for FPGA accelerator DSE, and evaluated a single accelerator design using high-level simulation. 
SECDA-DSE combines a structured DSE Explorer, responsible for generating and evaluating candidate hardware accelerator configurations, with an LLM Stack that performs reasoning-guided refinement using Retrieval-Augmented Generation (RAG), Chain-of-Thought (CoT) prompting, and parameter-efficient fine-tuning. The framework incorporates a feedback-driven evaluation loop in which simulation and hardware execution metrics are collected and reused to iteratively improve subsequent decisions, as shown in Figure~\ref{fig:SECDA-DSE}.

In this paper, we present SECDA-DSE and extend its evaluation by generating multiple accelerator kernels and performing end-to-end real hardware execution on an FPGA. 
Specifically, we evaluate generated accelerators for element-wise vector multiplication, 2D-convolution, and matrix transpose. The generated designs successfully execute on FPGA hardware and exhibit workload-specific architectural characteristics and resource utilization patterns, highlighting the ability of the framework to adapt exploration decisions across different computational workloads without extensive fine-tuning.

This paper makes the following contributions:

\begin{itemize}
    \item We present SECDA-DSE, an LLM-guided framework for FPGA accelerator DSE that combines structured exploration, RAG, CoT prompting, fine-tuning, and iterative hardware evaluation.

    \item We extend the evaluation of SECDA-DSE from a single simulation-based accelerator design to multiple accelerator kernels validated through end-to-end FPGA execution, including vector multiplication, convolution, and transpose kernels. All generated accelerators passed functional validation on the FPGA platform, confirming their execution correctness. Representative execution results show measured latencies of 154 ms for vector multiplication, 163 ms for 2D convolution, and 238 ms for transpose. The generated designs also exhibit different resource profiles, such as 21.82\% DSP utilization for vector multiplication and 8.85\% LUT utilization for transpose.

\end{itemize}

\section{Background and Related Work}
\label{background}

\subsection{FPGA Accelerator Co-Design}

FPGA-based accelerators have become an increasingly important platform for deploying modern AI workloads due to their flexibility, reconfigurability, and ability to provide energy-efficient computation across diverse applications~\cite{jiang2025fpga}. However, developing custom FPGA accelerators remains a complex process involving hardware-software co-design, hardware verification, high-level synthesis (HLS), and iterative performance optimization~\cite{gibsonDLAS2025}.

To shorten this accelerator design process, the SECDA methodology~\cite{haris2021secda} was introduced to simplify accelerator development by enabling rapid hardware-software co-design using SystemC-based simulation and FPGA execution flows. 
The SECDA-TFLite~\cite{haris2023secda} toolkit further extended this ecosystem by enabling TensorFlow Lite-based AI accelerator development and deployment using reusable architectural templates and automated hardware generation flows. More recently, SECDA-LLM~\cite{haris2024designing} explored the integration of Large Language Models (LLMs) within the SECDA ecosystem to assist accelerator development workflows.

While these frameworks significantly reduce development overhead and improve accessibility of FPGA accelerator design, the exploration of accelerator configurations and hardware optimization strategies still largely depends on manual DSE, requiring expert knowledge to reason over architectural and hardware constraints.


\subsection{DSE for FPGA Accelerators}

DSE is a widely used approach for identifying efficient hardware configurations under workload and device-specific constraints. Existing DSE methodologies for FPGAs commonly rely on exhaustive parameter sweeps, heuristic search methods, bayesian optimization, or techniques such as genetic algorithms to explore architectural parameter spaces~\cite{saeedi2024survey,biscontini2024machine}.

These approaches have proven their effectiveness in optimizing metrics such as latency, throughput, power consumption, and resource utilization. However, traditional DSE methods primarily operate over structured parameter spaces and often lack the ability to reason semantically about workload behavior, architectural patterns, or implementation trade-offs. Also, as the complexity of the accelerator increases, exhaustive exploration becomes computationally expensive due to the growing number of design permutations and hardware evaluation cycles~\cite{xu2025hardware}.

SECDA-DSE~\cite{sharma2026llm} provides an approach in which structured parameter exploration is augmented with an LLM-guided reasoning layer that can use prior hardware knowledge, retrieved design context, and evaluation feedback to refine accelerator configurations.


\subsection{LLMs for Hardware and Systems Design}


Recent advances in LLMs have motivated their use as assistants for engineering tasks that combine code generation, tool interaction, and structured decision making.
These capabilities have motivated growing interest in applying LLMs to hardware and system design workflows, including RTL generation, HLS assistance, compiler optimization, and architecture exploration~\cite{liao2024llms, li2025idse, zhang2026luminallmguidedgpuarchitecture}.

Several recent works have explored the use of LLMs for generating hardware descriptions or assisting accelerator development~\cite{fu2023gpt4aigchip, firouzi2024llm, vungarala2025sa}. Approaches such as RAG and CoT prompting also improve reasoning quality and contextual understanding for complex engineering tasks. In parallel, parameter-efficient fine-tuning techniques such as Low-Rank Adaptation (LoRA)~\cite{hu2022lora} enable adaptation of LLMs without requiring full model retraining.

Despite these advances, existing work largely focuses on isolated code generation tasks or standalone hardware synthesis assistance. Limited work has explored the integration of LLM-guided reasoning within structured FPGA DSE workflows that incorporate a retrieval-based context, iterative evaluation feedback, and real hardware execution.

In contrast, SECDA-DSE integrates structured design exploration, grounded retrieval reasoning, evaluation feedback, and hardware-aware refinement within a unified accelerator co-design framework. By combining SECDA-based hardware evaluation with LLM-guided reasoning, SECDA-DSE move towards adaptive and automated accelerator design workflows.

\section{SECDA-DSE}
\label{method}

SECDA-DSE extends the SECDA ecosystem by introducing an LLM-guided framework for accelerator DSE. The framework combines a structured DSE with reasoning-guided refinement to iteratively generate, evaluate, and improve accelerator configurations for target workloads and FPGA platforms.

\begin{figure}[!tb]
  \centering
  \includegraphics[width=0.95\linewidth]{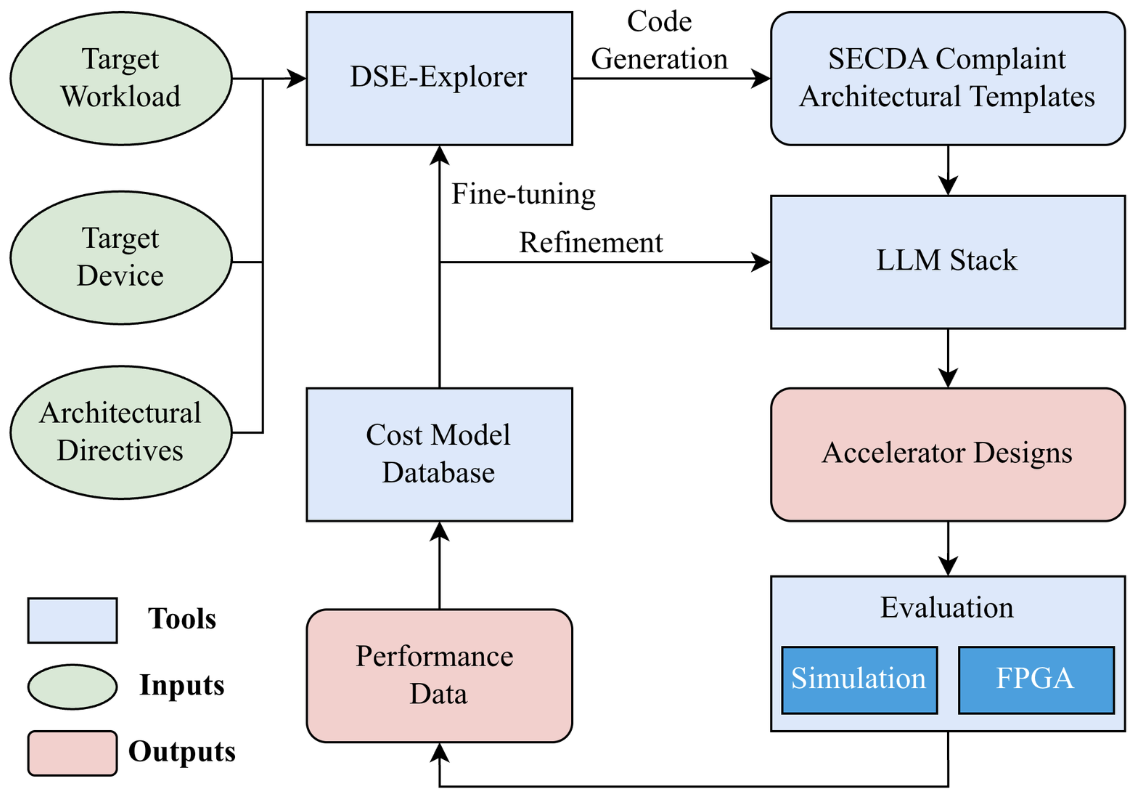}
  \caption{SECDA-DSE workflow showing the interaction between the DSE Explorer, LLM Stack, and Evaluation Module.}
  \label{fig:SECDA-DSE}
\end{figure}

The SECDA-DSE workflow is shown in Figure~\ref{fig:SECDA-DSE}. It takes as input: (i) a target workload (e.g., CNNs, DNNs); (ii) a target FPGA device; and (iii) architectural directives that constrain or guide architecture exploration, such as tiling parameters, compute parallelism, and memory hierarchy organization. Based on these inputs, SECDA-DSE generates SECDA native accelerator designs, evaluates them through simulation and hardware execution flows, and stores the resulting hardware datapoints for future refinement iterations.

Unlike unconstrained hardware generation approaches, SECDA-DSE operates within the SECDA ecosystem using template-constrained accelerator generation.
While constrained limits exploration to SECDA-compliant architectural patterns, it enables the reliable generation of synthesizable and executable accelerators while still allowing the LLM Stack to explore a large design space encompassing compute parallelism, memory organization, and dataflow configurations. 
Furthermore, the framework iteratively refines generated designs by reasoning about errors identified in earlier iterations.



\subsection{DSE Explorer}

The DSE Explorer is responsible for structured exploration of the accelerator design space. Its primary role is to generate candidate accelerator configurations under workload and FPGA-specific constraints. 

The DSE Explorer operates by generating permutations of architectural parameters, including compute unit dimensions, tiling strategies, buffer allocations, and dataflow configurations. These parameter sets are instantiated into SECDA-compliant accelerator templates, enabling integration with existing SECDA simulation and FPGA execution flows. 
Each generated accelerator configuration produces an individual design run consisting of the accelerator description, generated source files, HLS outputs, and FPGA execution artifacts. The resulting hardware design is evaluated using SECDA-based simulation and downstream synthesis flows to collect performance metrics including execution latency, resource utilization, and data movement costs.

These evaluation outputs are stored as hardware datapoints that are later re-used to fine-tune the LLM Stack and guide subsequent refinement iterations. By combining structured parameter exploration with evaluation-driven feedback, the DSE Explorer enables SECDA-DSE to navigate large accelerator design spaces more efficiently than exhaustive exploration approaches.


\subsection{LLM Stack}

\begin{figure}[!tb]
  \centering
  \includegraphics[width=0.95\linewidth]{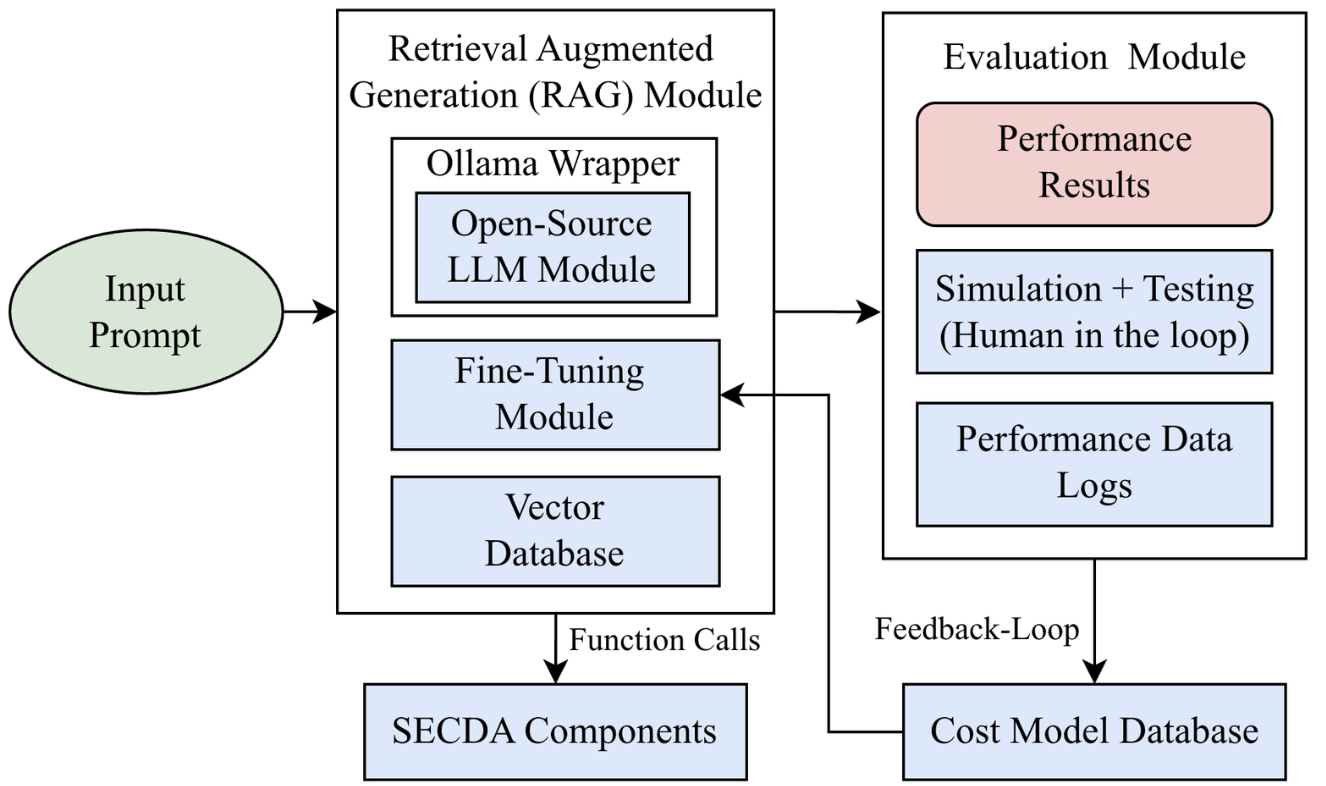}
  \caption{Overview of LLM Stack Architecture.}
  \label{fig:LLM_STACK}
\end{figure}

The LLM Stack acts as the reasoning and orchestration layer within SECDA-DSE. Its role is to analyze the generated accelerator configurations, reason about hardware trade-offs, and guide exploration refinement across iterations. 
It incorporates RAG, CoT prompting, and parameter-efficient fine-tuning to provide workload-aware accelerator design generation. Figure~\ref{fig:LLM_STACK} provides a high-level overview of the complete LLM Stack and interaction between components and across the modules.

\subsubsection{Retrieval-Augmented Generation}
The RAG module enables the LLM Stack to retrieve contextual information from a vectorized SECDA knowledge-base consisting of SECDA-TFLite source code, architectural templates, API-level context, and previously evaluated hardware datapoints. 
Rather than exposing the complete SECDA codebase or full hardware logs during each iteration, the RAG module retrieves only the most relevant code fragments and hardware summaries associated with the current exploration step. We employ a graph-based retrieval approach to efficiently identify relevant code context, using fuzzy matching on code comments to guide navigation across graph nodes. This method reduced retrieval overhead compared to standard nearest-neighbor retrieval approaches while improving contextual relevance during generation. 

This retrieval-based approach enables SECDA-DSE to maintain manageable context sizes while still providing sufficient architectural context for refinement. The retrieved information includes workload characteristics, prior accelerator configurations, hardware utilization summaries, and evaluation outcomes, before recommending new design configurations.

\subsubsection{Chain-of-Thought Prompting and Fine-Tuning}
SECDA-DSE employs CoT prompting to encourage structured multi-step reasoning during accelerator exploration. CoT prompts enable the LLM Stack to reason about workload behavior, hardware constraints, memory organization, and compute parallelism before generating candidate architectural refinements.

For model adaptation, SECDA-DSE uses parameter-efficient fine-tuning with LoRA~\cite{hu2022lora}. The fine-tuning dataset is built using previously explored accelerator configurations and their associated hardware evaluation outcomes generated via DSE-Explorer, with a human-in-the-loop for ensuring correctness.

Each hardware datapoint consists of workload and FPGA context, generated architectural configuration, simulation and execution success status, execution latency, and resource utilization metrics. These datapoints enable the LLM Stack to iteratively adapt toward generating configurations that are feasible, performance-aware, and target-device-aware.


\subsection{Evaluation and Feedback Loop}

The Evaluation Module is responsible for validating generated accelerator designs and collecting feedback used for iterative refinement. The workflow begins with a SECDA-based SystemC simulation to verify functional correctness and collect early-stage performance estimates. Generated accelerator designs are then passed through HLS and once succeed go through FPGA execution flows. 
The collected evaluation metrics include execution latency, FPGA resource utilization, DMA transfer characteristics, data movement costs, and hardware execution validation. These results are stored within the SECDA-DSE hardware datapoint database and reused during subsequent exploration iterations.

To reduce invalid accelerator proposals, SECDA-DSE constrains generation using SECDA native architectural templates and device-aware parameter ranges rather than relying on unconstrained free-form hardware generation. Candidate designs that fail simulation, violate resource constraints, or fail downstream synthesis flows are rejected and logged as negative datapoints, which are again fed back as a negative reinforcement to the LLM to reason and correct.

The iterative interaction between the DSE Explorer, LLM Stack, and Evaluation Module enables SECDA-DSE to repeatedly refine accelerator exploration decisions while leveraging prior hardware evaluation knowledge across workloads and FPGA platforms. 


\section{Evaluation}
\label{04_Results&Evaluation}

To extend the evaluation of SECDA-DSE, we generated and evaluated three accelerator workloads: element-wise vector multiplication, 2D convolution, and matrix transpose. These workloads were intentionally selected to represent different accelerator execution characteristics, including arithmetic intensive computation, convolution-oriented spatial compute, and memory movement dominated execution patterns. The prompts used for generating these accelerator designs are provided in the Appendix. 

The SECDA-DSE workflow begins with a natural language accelerator specification provided by the user. This prompt is processed through the LLM Stack, where the RAG module retrieves the relevant SECDA implementation context, architectural templates, and previously collected hardware datapoints from the vectorized knowledge-base. The retrieved context is then combined with CoT prompting to generate a structured accelerator generation prompt enriched with SECDA-specific architectural guidance.

Using this enriched context, SECDA-DSE generates native SystemC accelerator designs together with the required SECDA integration files. In the current implementation, the generated designs are passed through a human-in-the-loop evaluation workflow using the SECDA toolkit. The generated accelerators are first validated using HLS, and the designs that successfully pass HLS are then evaluated through downstream logic synthesis and FPGA execution flows.

The evaluation was performed using Vivado HLS 2019.2~\cite{vivadohls2019_2} targeting a Xilinx Zynq-7000 FPGA platform (\texttt{xc7z020-clg400-1}). The LLM Stack used TinyLlama~\cite{zhang2024tinyllama}, a lightweight 1.1B parameter model deployed locally through Ollama~\cite{ollama}. Initially, the LLM was only fine-tuned using hardware datapoints generated from matrix addition and matrix multiplication accelerator workloads.

Table~\ref{tab:workload_comparison} summarizes the hardware execution and FPGA utilization characteristics for the generated workloads. The primary objective of this evaluation was to validate SECDA-DSE across multiple accelerator workloads through the complete hardware generation workflow, including HLS, logic synthesis, and FPGA execution. 
The correctness of the generated accelerator designs is validated by performing an element-wise comparison between the outputs produced by the FPGA and reference results generated by a CPU-based software implementation.
At this stage, the focus is on end-to-end hardware execution rather than producing fully optimized accelerator implementations. As discussed in more detail in Section~\ref{Limitation and future work}, subsequent iterations of SECDA-DSE will focus on performance-optimized accelerator designs.

\begin{scriptsize}
\begin{table}[t]
\centering
\caption{FPGA execution and resource utilization comparison for generated accelerator designs.}
\label{tab:workload_comparison}
\setlength{\tabcolsep}{3pt}
\renewcommand{\arraystretch}{1.1}
\begin{tabular}{|l|c|c|c|}
\hline
\textbf{Metric} & \textbf{2D Conv} & \textbf{VMUL} & \textbf{Transpose} \\
\hline
\hline
Validation & PASSED & PASSED & PASSED \\
\hline
Latency (ms) & 163 & 135 & 238 \\
\hline
HWC cycles (1/2/3) & 1251/76/1250 & 52/26/51 & 393/311/387 \\
\hline
\hline
DMA recv size (bytes) & 256 & 64 & 512 \\
\hline
DMA send size (bytes) & 268 & 136 & 524 \\
\hline
DMA recv speed (MB/s) & 17.07 & 4.57 & 13.47 \\
\hline
DMA send speed (MB/s) & 38.29 & 19.43 & 30.82 \\
\hline
DMA recv wait (ms) & 15 & 14 & 38 \\
\hline
DMA send wait (ms) & 7 & 7 & 17 \\
\hline
\hline
BRAM utilization (\%) & 2.50 & 3.57 & 2.86 \\
\hline
DSP utilization (\%) & 1.36 & 21.82 & 5.45 \\
\hline
LUT utilization (\%) & 6.64 & 8.23 & 8.85 \\
\hline
FF utilization (\%) & 4.30 & 6.02 & 5.84 \\
\hline
\end{tabular}
\end{table}
\end{scriptsize}

Despite being generated from natural language specifications, the produced accelerator designs were executable without requiring manual modification of the generated accelerator logic. The generated workloads exhibited distinct hardware characteristics depending on the computational structure of the workload. For example, vector multiplication workload emphasized arithmetic execution and parallel compute behavior, resulting in higher DSP utilization (\(21.82\%\)) compared to the convolution (\(1.36\%\)) and transpose (\(5.45\%\)) accelerators. While the transpose workload demonstrated a more memory movement dominated execution pattern with increased DMA transfer sizes and higher FPGA logic utilization (\(8.85\%\) LUT usage).

Using hardware counter (HWC) cycles and DMA profiling metrics, SECDA-DSE identifies bottlenecks within generated accelerator designs and uses this information during iterative refinement. HWC1 measures data loading wait cycles, HWC2 captures computation time, and HWC3 records write-back cycles. 
The 2D convolution accelerator demonstrated substantially higher cycle counts (\(1251/76/1250\)) compared to the vector multiplication (\(52/26/51\)) due to the increased buffering and data reuse complexity, while the transpose accelerator showed intermediate execution complexity (\(393/311/387\)) because of runtime matrix reorganization overheads. 
Similarly, the transpose workload required the largest DMA transfers (\(512\) receive bytes and \(524\) send bytes), reflecting its memory-centric execution pattern, whereas vector multiplication emphasized arithmetic throughput with smaller transfers. These profiling metrics are stored as hardware datapoints and reused by the LLM Stack during subsequent refinement iterations.

SECDA-DSE generated these accelerator designs using a relatively small 1.1B parameter model with only limited initial fine-tuning datapoints. Despite the constrained training setup, the framework successfully generated FPGA executable accelerator designs across multiple workload types while maintaining SECDA compatibility and valid FPGA execution flows.

The three generated accelerator workloads also exhibited different convergence behaviors during iterative refinement. The vector multiplication required four iterative refinement cycles before producing a valid FPGA executable design. 
In contrast, the 2D convolution accelerator successfully generated a valid design in a single generation cycle. The transpose accelerator required nine refinement cycles overall, where the first two iterations produced an HLS synthesizable design while an additional seven iterations were required to successfully complete downstream logic synthesis and FPGA execution. Throughout these iterations, human intervention was restricted to operational tasks required by the SECDA validation workflow, such as transferring generated source files, verifying execution paths and running evaluation scripts to collect hardware metrics. The generated accelerator designs were not manually modified at any stage.


These results provide early evidence that combining structured DSE, retrieval-grounded reasoning, and iterative hardware evaluation can enable adaptive accelerator generation workflows capable of handling diverse workload characteristics while reducing the manual effort traditionally required during FPGA accelerator development.

\section{Limitations and Future Work}
\label{Limitation and future work}

While SECDA-DSE demonstrates promising early-stage results across multiple accelerator workloads, the framework is still in an active development phase and currently presents some limitations. 
One of the primary challenges is that a sufficiently diverse set of hardware datapoints are initially required for effective fine-tuning and refinement of the LLM Stack. 
Furthermore, the current workflow still incorporates a human-in-the-loop evaluation stage for validating generated designs before FPGA execution. Although this improves reliability during early-stage development, scaling the framework toward larger workload coverage and broader exploration spaces will require significantly higher levels of automation. 
Another limitation is that the current evaluation covers only a small set of workloads on a single FPGA platform. Although the generated accelerators demonstrate workload-aware behavior and successful FPGA execution, broader evaluations across more workloads, architectures, and FPGA families are needed to assess the generalization capability of SECDA-DSE.

In future work, our goal is to fully automate the accelerator evaluation workflow by integrating the SECDA toolkit directly into SECDA-DSE through a model context protocol (MCP) server. This integration will allow the LLM Stack to directly invoke SECDA testing, simulation, HLS, and hardware execution functions without requiring manual intervention, enabling a more autonomous hardware exploration workflow. 
Additionally, the current evaluation focuses on a relatively small set of workloads and a single FPGA platform, as future work we plan to extend evaluation of SECDA-DSE across multiple parameter scales, LLMs and FPGAs to evaluate how model size, architecture and choice of the FPGA board influence accelerator generation quality and hardware exploration capability.

\section{Conclusion}
\label{Conclusion}

We presented SECDA-DSE, an LLM-guided framework for FPGA accelerator design space exploration built upon the SECDA ecosystem, and extended its evaluation through end-to-end FPGA execution. SECDA-DSE combines structured DSE, retrieval-augmented reasoning, Chain-of-Thought prompting, and iterative hardware evaluation to generate accelerator designs from natural language specifications. 
Specifically, we generated and executed element-wise vector multiplication, 2D convolution, and matrix transpose accelerators targeting a Xilinx Zynq-7000 FPGA platform. The generated designs successfully progressed through the FPGA implementation flow and demonstrated workload-aware hardware behavior across different compute and memory access patterns. 
Our results provide further evidence that combining LLM-guided reasoning with structured FPGA design workflows can support adaptive accelerator generation across diverse workloads while reducing manual design effort and advancing more automated hardware design methodologies. The successful FPGA execution of all generated accelerators further demonstrates the ability of SECDA-DSE to adapt across different workload characteristics.


\section*{Acknowledgements}

All authors gratefully acknowledge UK Research and Innovation (UKRI) and Engineering and Physical Sciences Research Council (EPSRC) funding for the AI Hub for Productive Research and Innovation in Electronics (APRIL) AI Hub [grant number EP/Y029763/1].



\balance

\bibliographystyle{IEEEtranS}
\bibliography{refs}

@article{mohaidat2024survey,
  title={{A Survey on Neural Network Hardware Accelerators}},
  author={Mohaidat, Tamador and Khalil, Kasem},
  journal={IEEE Transactions on Artificial Intelligence},
  volume={5},
  number={8},
  pages={3801--3822},
  year={2024},
  publisher={IEEE}
}

@misc{sharma2026llm,
  author={Sharma, Vinamra and Fu, Xingjian and Haris, Jude and Cano, Jos{\'e}},
  title={{LLM-Driven Design Space Exploration of FPGA-based Accelerators}},  
  year={2026},  
  note={arXiv:2401.12345},
}

@inproceedings{haris2021secda,
  title={{SECDA: Efficient Hardware/Software Co-Design of FPGA-based DNN Accelerators for Edge Inference}},
  author={Haris, Jude and Gibson, Perry and Cano, Jos{\'e} and Agostini, Nicolas Bohm and Kaeli, David},
  booktitle={IEEE 33rd Int. Symposium on Computer Architecture and High Performance Computing (SBAC-PAD)},  
  year={2021},  
}

@article{haris2023secda,
  title={{SECDA-TFLite: A toolkit for efficient development of FPGA-based DNN accelerators for edge inference}},
  author={Haris, Jude and Gibson, Perry and Cano, Jos{\'e} and Agostini, Nicolas Bohm and Kaeli, David},
  journal={Journal of Parallel and Distributed Computing},
  volume={173},
  pages={140--151},
  year={2023},
  publisher={Elsevier}
}

@misc{haris2024designing,
  title={{Designing Efficient LLM Accelerators for Edge Devices}},
  author={Haris, Jude and Saha, Rappy and Hu, Wenhao and Cano, Jos{\'e}},  
  note={arXiv:2408.00462},
  year={2024}
}

@article{saeedi2024survey,
  title={{A Survey on Design Space Exploration Approaches for Approximate Computing Systems}},
  author={Saeedi, Sepide and Piri, Ali and Deveautour, Bastien and O’connor, Ian and Bosio, Alberto and Savino, Alessandro and Di Carlo, Stefano},
  journal={Electronics},
  volume={13},
  number={22},
  pages={4442},
  year={2024},
  publisher={MDPI}
}

@article{biscontini2024machine,
  title={{Machine Learning for FPGA Electronic Design Automation}},
  author={Biscontini, Armando and Popovici, E and Temko, Andriy},
  journal={IEEE Access},
  volume={12},
  pages={182640--182662},
  year={2024},
  publisher={IEEE}
}

@inproceedings{liao2024llms,
  title={{Are LLMs Any Good for High-Level Synthesis?}},
  author={Liao, Yuchao and Adegbija, Tosiron and Lysecky, Roman},
  booktitle={Proceedings of the 43rd IEEE/ACM International Conference on Computer-Aided Design},
  pages={1--8},
  year={2024}
}

@misc{li2025idse,
  title={{iDSE: Navigating Design Space Exploration in High-Level Synthesis Using LLMs}},
  author={Li, Runkai and Xiong, Jia and Wang, Xi},  
  note={arXiv:2505.22086},
  year={2025}
}

@misc{zhang2026luminallmguidedgpuarchitecture,
      title={{LUMINA: LLM-Guided GPU Architecture Exploration via Bottleneck Analysis}}, 
      author={Tao Zhang and Rui Ma and Shuotao Xu and Peng Cheng and Yongqiang Xiong},
      year={2026},    
      note={arXiv:2603.05904},
}

@inproceedings{fu2023gpt4aigchip,
  title={{GPT4AIGChip: Towards Next-Generation AI Accelerator Design Automation via Large Language Models}},
  author={Fu, Yonggan and Zhang, Yongan and Yu, Zhongzhi and Li, Sixu and Ye, Zhifan and Li, Chaojian and Wan, Cheng and Lin, Yingyan Celine},
  booktitle={2023 IEEE/ACM International Conference on Computer Aided Design (ICCAD)},  
  year={2023},  
}

@inproceedings{firouzi2024llm,
  title={{LLM-AID: Leveraging Large Language Models for Rapid Domain-Specific Accelerator Development}},
  author={Firouzi, Farshad and Nakkilla, Sri Sai Rakesh and Fu, Chenghao and Banerjee, Sanmitra and Talukdar, Jonti and Chakrabarty, Krishnendu},
  booktitle={43rd IEEE/ACM International Conference on Computer-Aided Design},
  pages={1--9},
  year={2024}
}

@inproceedings{vungarala2025sa,
  title={{SA-DS: A Dataset for Large Language Model-Driven AI Accelerator Design Generation}},
  author={Vungarala, Deepak and Nazzal, Mahmoud and Morsali, Mehrdad and Zhang, Chao and Ghosh, Arnob and Khreishah, Abdallah and Angizi, Shaahin},
  booktitle={2025 IEEE International Symposium on Circuits and Systems (ISCAS)},
  pages={1--4},
  year={2025},
}

@article{hu2022lora,
  title={{LoRA: Low-Rank Adaptation of Large Language Models}},
  author={Hu, Edward J and Shen, Yelong and Wallis, Phillip and Allen-Zhu, Zeyuan and Li, Yuanzhi and Wang, Shean and Wang, Liang and Chen, Weizhu and others},
  journal={ICLR},
  volume={1},
  number={2},
  pages={3},
  year={2022}
}

@misc{zhang2024tinyllama,
  title={{TinyLlama: An Open-Source Small Language Model}},
  author={Zhang, Peiyuan and Zeng, Guangtao and Wang, Tianduo and Lu, Wei},  
  year={2024},
  note={arXiv:2401.02385},
}

@misc{ollama,
  author = {{Ollama}},
  title  = {Ollama: Run Large Language Models Locally},
  year   = {2026},
  note   = {[Online]. Available: https://ollama.com. Accessed: Jun. 5, 2026}
}

@misc{vivadohls2019_2,
  title        = {Vivado High-Level Synthesis},
  author       = {{Xilinx, Inc.}},
  year         = {2019},
  version      = {2019.2}
}

@article{hou2024large,
  title={{Large Language Models for Software Engineering: A Systematic Literature Review}},
  author={Hou, Xinyi and Zhao, Yanjie and Liu, Yue and Yang, Zhou and Wang, Kailong and Li, Li and Luo, Xiapu and Lo, David and Grundy, John and Wang, Haoyu},
  journal={ACM Transactions on Software Engineering and Methodology},
  volume={33},
  number={8},
  pages={1--79},
  year={2024},
  publisher={ACM New York, NY}
}

@inproceedings{fan2023large,
  title={{Large Language Models for Software Engineering: Survey and Open Problems}},
  author={Fan, Angela and Gokkaya, Beliz and Harman, Mark and Lyubarskiy, Mitya and Sengupta, Shubho and Yoo, Shin and Zhang, Jie M},
  booktitle={2023 IEEE/ACM International Conference on Software Engineering: Future of Software Engineering (ICSE-FoSE)},
  pages={31--53},
  year={2023},
  organization={IEEE}
}

@misc{jiang2025fpga,
  title={{FPGA-based Acceleration for Convolutional Neural Networks: A Comprehensive Review}},
  author={Jiang, Junye and Zhou, Yaan and Gong, Yuanhao and Yuan, Haoxuan and Liu, Shuanglong},  
  year={2025},
  note={arXiv:2505.13461},
}

@misc{xu2025hardware,
  title={{Hardware Acceleration for Neural Networks: A Comprehensive Survey}},
  author={Xu, Bin and Banerjee, Ayan and Gupta, Sandeep},  
  year={2025},
  note={arXiv:2512.23914}
}

@article{gibsonDLAS2025,
  title      = {{{DLAS}}: {{A Conceptual Model}} for {{Across-Stack Deep Learning Acceleration}}},
  shorttitle = {{{DLAS}}},
  author     = {Gibson, Perry and Cano, Jose and Crowley, Elliot and Storkey, Amos and O’boyle, Michael},
  year       = {2025},
  journal    = {ACM Transactions on Architecture and Code Optimization (TACO)}
}

\appendix

\section{Appendix}


This appendix provides the initial natural language prompts used for generating the evaluated accelerator workloads presented in Section~\ref{04_Results&Evaluation}.


\subsection{Element-wise Vector Multiplication}
\emph{
I would like to create a hardware accelerator design. The accelerator should be able to take two input vectors: X and Y, both of length L.
The accelerator should perform an element-wise multiplication
operation and produce an output vector Z.
The accelerator has two AXI-Stream based interfaces for loading
X and Y data into custom X and Y buffers.
The accelerator should also have a fixed length parameter L.
Once the data is loaded, the accelerator should execute the
element-wise multiplication in parallel and store the results
in buffer Z within the compute module.
The loading should be performed using a load module. Finally, the results should be written back to main memory using a store module that outputs via an AXI-Stream interface.
Create the accelerator description using SystemC and SECDA. The compute module should be capable of performing L operations in parallel.
}

\subsection{2D Convolution}
\emph{
I would like to create a hardware accelerator design.
The accelerator should be able to take two input matrix: one for weights and another for inputs.
The weights have the dimensions in follow data order: input channels (IC), output channels (OC), kernel width (KW) and then kernel height (KH) .
The input matrix has the dimensions in follow data order: input channels (IC), input width (IW) and then input height (IH).
The output matrix has the dimensions in follow data order: output channels (OC), output width (OW) and then output height (OH).
We assume padding is 0 and stride is 1 and dilation is 1 for the convolution operation.
So the output width (OW) and output height (OH) can be calculated as follows:
OW = IW - KW + 1
OH = IH - KH + 1
The accelerator should perform convolution operation and produce a output matrix Z.
The accelerator has two AXI-Stream based interfaces for loading weights and inputs into custom weights and inputs buffers.
All dimension are fixed and defined as parameters in the accelerator design.
Once the data is loaded, the accelerator should be abe to execute the CONV operation in parrallel and save it in buffer Z, within the compute module.
The loading should be done with load module.
And finally the storing of the results back to main memory should be done in the store module which writes to an output AXI-Stream.
Create the accelerator description using SystemC and SECDA.
}

\subsection{Matrix Transpose}
\emph{
I would like to create a hardware accelerator design.
The accelerator should be able to take a input matrix X of dimensions (M, N), transpose it into an output matrix Z of dimensions (N, M).
Where M, N are not fixed and are provided as data at runtime, before matrix X is loaded into the accelerator. The accelerator should be able to handle any dimensions of M and N, as long as they fit within the memory constraints (buffer size) of the accelerator design.
The accelerator has an AXI-Stream based interface for loading the input matrix X into a custom input buffer.
Once the data is loaded, the accelerator should be able to execute the transpose operation and save the result in buffer Z, within the compute module.
The loading should be done with load module.
The Compute module should be able to perform the transpose operation. It should d
And finally the storing of the results back to main memory should be done in the store module which writes to an output AXI-Stream..
Create the accelerator description using SystemC and SECDA.
}

\end{document}